\documentstyle[aps,preprint]{revtex}
\begin{document}
\draft
\title { MACROSCOPIC FEATURES OF LIGHT HEAVY-ION FISSION REACTIONS }
\author { C. Beck }
\address{\it Centre de Recherches Nucl\'eaires, Institut National de Physique
Nucl\'eaire et de Physique des Particules - Centre National de la Recherche
Scientifique/Universit\'e Louis Pasteur,B.P.28, F-67037 Strasbourg Cedex 2,
France }
\author { A. Szanto de Toledo }
\address {\it Instituto da Universidade de S\~ao Paulo, Departemento de
Fisica Nuclear, Caixa Postal 66318 - 05389-970 - S\~ao Paulo, Brazil}

\date{\today}
\maketitle

\begin{abstract}
{ Global macroscopic features observed in the fully-damped binary processes in
light di-nuclear systems, such as limiting angular momenta, mean total kinetic
energies and energy thresholds for fusion-fission processes (''fission
thresholds") are presented. Their deduced systematics are consistent with that
obtained for heavier systems and follow a fusion-fission picture which can be
described by a realistic rotating liquid drop model considering diffuse-surface
and finite-nuclear-range effects.}
\end{abstract}


In recent years, light heavy-ion collisions have been extensively studied over
a wide range of low bombarding energies (E$_{lab}$ $\leq$ 10 MeV/nucleon) for
various target+projectile combinations, thus well establishing that the
fusion-fission process (FF) in {\bf light} di-nuclear systems$^{1-7}$ (in the A
$\leq$ 60 mass region) has to be taken into account when exploring the
limitations of the complete fusion (CF) process at high excitation energies and
large angular momenta. The main extracted properties of the fully energy damped
processes for these light heavy-ion reactions have been well explained with
several different statistical descriptions (based for example either on the
saddle point picture$^{5}$ or on the scission point formalism$^{4}$),
indicating that FF is significant in this mass region. In particular FF
calculations$^{4-7}$ have been found successful to describe experimental
observables, such as mass, charge and energy distributions of the fully-damped
yields of most if not all the studied systems. However a more systematic
understanding of the total fusion process is still needed as soon as the FF
yield is found non negligeable due to a lowering of the fission barriers. By
using the basic hypothesis of the rotating liquid drop model$^{8}$ (LDM),
diffuse-surface$^{9}$ and finite-nuclear-range effects have been found to be of
crucial importance$^{5,10,11}$ to account for experimental FF cross sections
within the framework of the early steady-state theory of Bohr and Wheeler. We
will show in this Brief Report that a global macroscopic description of the
main features of FF needs also these effects to be incorporated for {\bf light}
nuclei.

In the so-called first regime of the fusion process (region I), near the
Coulomb barrier, the fusion cross section is determined by the barrier
penetration probability and, for incident energies up to around twice the
Coulomb barrier energy, the reaction cross section is mainly dominated by the
CF yield. At higher incident energies (in region II), the CF yield tends to
saturate due to the activation of other competing mechanisms such as deep
inelastic (DI) collisions. Heavy-ion resonances and DI orbiting processes have
also been shown to occur for partially damped and large near-grazing partial
wave components of the incident flux$^{12}$ at incident energies higher than
the bending energy where the CF cross section begins to saturate. The DI
processes can lead the two interacting nuclei to form a di-nuclear complex,
from which they can either fuse into a compound nucleus (CN) after complete
equilibration of all the degrees of freedom, or escape from the ion-ion
potential well, producing a damped orbiting process. The interplay between the
nuclear potential and the friction forces, which control the damping of both
the kinetic energy and the angular momentum, will determine whether fusion or
faster binary decay processes may occur. It has been also shown that the degree
of competition between these two classes of processes is correlated to the
number of open channels (NOC) available for the decay of the di-nuclear
system$^{12,13}$. Large NOC tend to favour the occurence of statistical
processes with regard to faster mechanisms which retain the memory of the
entrance channel in sharp contrast to the theory of Bohr and Wheeler.

In region III, at higher bombarding energies, the survival probabilities of
both the target and the projectile are drastically suppressed during the
collision and consequently the incomplete fusion components become significant.
In this energy regime the measured CF cross sections show a rapid decrease
which corresponds to an experimental limiting angular momentum. The compiled
limiting angular momenta have been taken from Ref.14 or extracted from more
recent CF excitation measurements measurements$^{15-20}$ which have all shown a
critical angular momentum limitation. These maximum angular momentum data, due
to a general instability of the composite system against fission, lie
systematically well below the LDM angular momentum limit$^{8}$ (dashed line)
for which the fission barrier vanishes. The data are found, however in
agreement with the predictions of a modified version of LDM$^{10}$ which
includes finite-range corrections of the nuclear interaction by means of a
Yukawa-plus-exponential attractive potential$^{9}$ and diffuse surface
effects$^{10}$. The finite-range corrections produce for each angular momentum
a lowering of the fission barrier due to the attractive forces between surfaces
of the two nascent fragments at the saddle point. The magnitude of the
reduction in fission barriers which increases with decreasing A is one of the
main reason why FF has been recently investigated for {\bf light} di-nuclear
systems. The vanishing fission barriers displayed in Fig.1 have been calculated
for the beta-stable nuclei (solid line) using the finite-nuclear-range LDM
(FRLDM) extended to rotating nuclei by Sierk$^{10}$. In this model$^{10}$ the
attractive force of finite-range was permitted between the nascent fragments at
the saddle point in order to reproduce the experimentally deduced
angular-momentum dependent barriers$^{10,11}$. The increasing discrepancy with
increasing mass number (in the vicinity of A $\leq$ 100 where the
Businaro-Gallone transition is expected to occur) is observed in Fig.1 and
interpreted by the possible occurence, in heavier systems$^{20}$, of
fast-fission and quasi-fission mechanisms (considered as an equivalent to a DI
orbiting process for lower mass number) competing with the statistical FF for
the highest partial waves. In this high-energy domain, the compound system is
formed at very high excitation energy and high angular momenta inducing quite
large deformation effects in the evaporative cascade. In addition to the
evaporation of neutrons, protons and alphas, the fused system in this region
may undergo binary decay through the emission of complex fragments and
intermediate mass fragments$^{21}$ and/or an asymmetric fission process$^{5}$.
Finally in the intermediate energy regime (E$_{lab}$ $\geq$ 40 MeV/nucleon) ,
the nuclear system before its complete dissassembly (multifragmentation) is
capable of reaching an upper limit for the temperature at which a
``vaporization mechanism" into light particles may occur.

Although the FF cross section in light heavy-ion reactions represents only a
small fraction (less than 5-10 $\%$) of the total reaction cross section,
several systems were investigated in detail$^{1-7}$ and characteristic
macroscopic features were clearly established$^{5}$. The statistical origin of
the observed fully-damped fragments has been recently established for several
light di-nuclei such as $^{19}$F and $^{20,21}$Ne (investigated by means of the
$^{9}$Be,$^{10,11}$B+$^{10}$B reactions respectively$^{15,16}$) $^{27,28}$Al
(investigated by means of the $^{16}$O+$^{11}$B, $^{17}$O+$^{10,11}$B
$^{18}$O+$^{10}$B and $^{19}$F+$^{9}$Be reactions respectively$^{7}$), $^{47}$V
(investigated by means of the $^{35}$Cl+$^{12}$C, $^{31}$P+$^{16}$O and
$^{23}$Na+$^{24}$Mg reactions$^{6}$) and $^{48}$Cr (investigated by means of
the $^{24}$Mg+$^{24}$Mg, $^{36}$Ar+$^{12}$C and $^{20}$Ne+$^{28}$Si
reactions$^{22}$) as populated by different entrance channels having very
different mass-asymmetries, thus verifying whether the Bohr hypothesis is
fulfilled or not. In other cases for which this has not been experimentally
possible, the experimental data have been compared to FF predictions$^{1,5}$.

Generally all of the observables, including the fragment total kinetic energies
(TKE), obtained in these experiments$^{1,5-7,22}$ are very well described by a
FF picture using the statistical model$^{4,5}$. We have
compiled in Fig.2 the mean TKE values, of the symmetric fission fragments
produced in light heavy-ion systems. The systematics due to Viola$^{23}$
(dashed lines), which predicts a linear dependence of TKE with the Coulomb
parameter Z$^{2}$/A$^{1/3}$, is capable of describing the data set available in
the literature$^{23,24}$, but fails in the case of low Z fissioning nuclei as
shown by the insert of Fig.2. Due to the diffuse nature of the nuclear surfaces
of light nuclei and the associated perturbations of the necking degree of
freedom consistent with both the droplet model$^{25}$ and FRLDM$^{10}$
calculations, a change in slope at low values of Z$^{2}$/A$^{1/3}$ (see insert
of Fig.2) leading to vanishing TKE values as Z approaches zero may be expected.
This effect is illustrated by the solid curves drawn in Fig.2 which has been
calculated by the following formula :

$$TKE = Z^{2}/(aA^{1/3}+bA^{-1/3}+cA^{-1})$$

where the values of the parameters are a = 9.65 MeV$^{-1}$, b = -58.1
MeV$^{-1}$ and c = 188 MeV$^{-1}$ respectively as a result of a simple weighted
least-squares fitting procedure. The mass dependence of this formula is
justified by the fact that for spherical nuclei the extension of the charge
distributions around their centers has this type of dependence as required by
the droplet model$^{25}$ to take the diffuseness into account in LDM$^{8}$.
Although second-order effects such as nuclear structure (mainly shell effects)
and pairing effects have not been taken into account, this new simple universal
expression of TKE, valid for a very wide range of fissioning systems extended
to the lighter ones, appears to be quite useful for its prediction
capabilities.

At this point it is interesting to note that all the data available for light
systems were obtained from FF yields. Although the TKE values were extracted at
the threshold energies for the process, a small contribution of the rotational
energy term is included. This contribution is expected to be not very
significant in cases where the effective moment of inertia in the
double-spheroid approximation of the saddle point is quite large and the fusion
critical angular momentum at the threshold is rather small. In this context,
one has to admit that some of the small discrepancies observed for medium
weighted
nuclei (200 $\leq$ Z$^{2}$/A$^{1/3}$ $\leq$ 500) can possibly be explained by
this additional rotational component present in the experimental points.
However, it is still relevant to do the comparison with heavier, more
fissile systems data as those previously compiled by Viola et al.$^{23}$.

The measurements of the excitation functions for the total fission cross
sections for the $^{10}$B+$^{10}$B (Ref.3), $^{18}$O+$^{10}$B (Ref.7),
$^{35}$Cl+$^{12}$C (Ref.4 and 24) and $^{16}$O+$^{40,44}$Ca (Ref.1) reactions
have shown that the FF cross sections rise rapidly with increasing bombarding
energies and then more slowly at higher energies. When plotted as a function of
1/E$_{cm}$ the excitation functions present a simple linear relation similar to
the case of CF processes in region I (see Ref.12 or 13 for instance). This
behaviour is a characteristic signature of a statistical CN emission and well
predicted by statistical model calculations$^{1-7}$. These calculations start
with the CN formation hypothesis and then follow the decay of the system by
first chance binary fission or light-particle emission and subsequent
light-particle or photon emission. Since the LDM$^{8}$ predicts too high
fission barriers, the mass-asymmetric fission barriers are calculated following
the procedure outlined in FRLDM$^{10}$ in order to incorporate diffuse-surface
and finite-nuclear-range effects. The transition-state method has been most
notably successful in accounting for many of the observed features of the
fission process in the {\bf light} di-nuclear systems$^{5}$ as well as for the
emission of complex fragments from heavier compound nuclei$^{21}$.

Based upon the available FF excitation function data (including the
fully-damped yields data for the $^{12}$C+$^{24}$Mg,$^{28}$Si and
$^{14}$N+$^{28}$Si reactions taken from Ref.26 and the fission data for the
$^{35}$Cl+$^{52}$Ni reaction$^{27}$), it has been possible to extract
experimental values of ``fission threshold" ($E^{threshold}_{fission}$) which
correspond to the intersection of the FF cross section curve with the
(1/E$_{cm}$) axis. The experimental data shown in Fig.3 indicate that
$E^{threshold}_{fission}$ has a linear relationship with the CN Coulomb
parameter Z$^{2}$/A$^{1/3}$. The reduced values of
$E^{threshold}_{fission}$/(Z$^{2}$/A$^{1/3}$) for different nuclei ranging from
mass 20 to mass 97, plotted as a function of x the LDM fissility
parameter$^{8}$ , appear to be pratically constant within the error bars.
Although there exists no simple relashionship with the FF macroscopic energies,
these ``fission thresholds" might be possibly associated to fission barriers
after consideration of the contribution of the centrifugal term (as shown for
the TKE systematics) and also that of the saddle point deformations which might
be non negligeable at these incident energies. The lack of any obvious
dependence from system to system is not yet well understood. This study calls
for new experimental works in the 0.35 $\leq$ x $\leq$ 0.50 fissility region
for which the Businaro-Gallone transition from asymmetrical to symmetrical
fission is expected to occur. In the meantime a more theoretical approach of
the fission barriers in this mass region will surely require more detailed
excitation function measurements for ``sub-threshold" bombarding energies in
order to check the ``universality" of the complex fragment emission as proposed
very recently by Moretto$^{21}$ for more massive systems. Experimental studies
are being currently undertaken.

In summary, a systematic examination of the main general characteristics of the
fusion-fission process (limiting angular momenta, mean total kinetic energies
and ``fission thresholds") in light heavy-ion reactions suggests that the
rotating liquid drop model can be extended for very {\bf light} nuclei when the
diffuse nature of nuclear surfaces and finite-nuclear-range effects have been
explicitly taken into account as previously proposed for heavier
nuclei$^{10,21}$. Although more refined theoretical studies will have to be
undertaken to investigate second order effects (such as nuclear structure
effects, proximity and/or viscosity effects, the temperature dependence of the
surface energy, ...), this systematic study suggests that a very crude
macroscopic picture of nuclear matter at high excitation energy and angular
momentum remains a reasonable way to describe the collective nature of hot
nuclei as light as the $^{19}$F and $^{20}$Ne nuclei which can be considered as
remnants of liquid droplets.

\bigskip

The authors would like to thank S.J. Sanders and R.M. Freeman for their
comments on the manuscript. This work has been partly supported by CNRS of
France. One of us (C.B.) would like to acknowledge CNPq of Brazil for an
additional financial support during his stay at the University of S\~ao Paulo
within the framework of a CNRS/CNPq collaboration program.


%
%
\begin{figure}
Fig.1 : Experimental (points) and calculated LDM$^{8}$ (dashed line) and
FRLDM$^{10}$ (solid line) limiting angular momenta for fusion as a function of
the CN mass number A$_{CN}$ of beta-stable nuclei. The solid points have been
previously compiled in Ref.14. The open points correspond to the following
reactions (compound systems): $^{10}$B+$^{9}$Be ($^{19}$F) and
$^{10}$B+$^{10,11}$B ($^{20,21}$Ne) taken from Refs.15-16,
$^{28}$Si+$^{12}$C,$^{40}$Ca ($^{40}$Ca,$^{68}$Se) from Refs.17-18,
$^{16}$O+$^{40}$Ca ($^{56}$Ni) from Ref.19 and, $^{40}$Ca+$^{40}$Ca ($^{80}$Zr)
from Ref.20.
\end{figure}

\begin{figure}
Fig.2 : Most probable mean TKE release in fission as a function of the Coulomb
parameter Z$^{2}$/A$^{1/3}$ of the fissioning nucleus. Open triangles have been
taken from previous existing compilations$^{23-24}$. Experimental solid points
have been compiled from the data given in Refs.3-7,15-16,19-20. The dashed
lines
are the result of the Viola systematics$^{23}$ whereas the solid lines are the
result of the fitting procedure discussed in the text.
\end{figure}

\begin{figure}
Fig.3 : Experimental ``fission thresholds" divided by the Coulomb parameter
$E^{threshold}_{fission}$/(Z$^{2}$/A$^{1/3}$) plotted as a function of the
fissility parameter x for the indicated reactions and compound systems.
\end{figure}

%
%

\end{document}